
\input harvmac

\def\CZ{{\cal Z}}

\def\CO{{\cal O}}
\def\l{\ell}
\def\CA{{\cal A}}
\def\CW{{\cal W}}

\def\p{\partial}

\def\R{\relax{\rm I\kern-.18em R}}
\font\cmss=cmss10 \font\cmsss=cmss10 at 7pt
\def\Z{\relax\ifmmode\mathchoice
{\hbox{\cmss Z\kern-.4em Z}}{\hbox{\cmss Z\kern-.4em Z}}
{\lower.9pt\hbox{\cmsss Z\kern-.4em Z}}
{\lower1.2pt\hbox{\cmsss Z\kern-.4em Z}}\else{\cmss Z\kern-.4em Z}\fi}
\def\pl{{\it  Phys. Lett.}}

\def\mpl{{\it Mod. Phys.   Lett.}}
\def\np{{\it Nucl. Phys. }}

    \lref\Imat{I. Kostov and M. Staudacher, \np B  384 (1992) 459}
 \lref\Iade{I. Kostov, \np B 326, (1989) 583}
\lref\ooo{I. Kostov, \pl\ B 238 (1990) 181}
 \lref\adem{I. Kostov, \pl  B 297 (1992)74}
  \lref\jjev{A. Jevicki and J. Rodrigues, preprint
LPTHENS-93-52, hep-th 9312118; N. Nakazawa, NBI preprint
NBI-HE-94-54}
\lref\kko{V. Kazakov and I. Kostov, \np B 386 (1992) 520}
\lref\Icar{I. Kostov, ``Strings embedded in Dynkin diagrams, in the  Proc.
of the Cargese meeting {\it Random Surfaces, Quantum Gravity and Strings},
Saclay Preprint SPhT/90-133}
\lref\djv{ A. Jevicki and B. Sakita, \np B165 (1980) 511}
\lref\Idis{I.K. Kostov,  \np B 376 (1992) 539}
 \lref\aff{J. Affleck,Nucl. Phys. B185 (1981) 346}
\lref\dvv{M. Fukuma, H. Kawai and R. Nakayama, {\it Int. J. Mod. Phys.} A
6 (1991) 1385;
R. Dijkgraaf, H. Verlinde and E. Verlinde, \np B 348 (1991) 435}
 \lref\onn{I. Kostov, \mpl A 4 (1989) 217, M. Gaudin and I. Kostov, \pl
B220 (1989) 200,
I. Kostov and M. Staudacher, \np B 384 (1992), B. Eynard and J.
Zinn-Justin, \np  B 386
(1992) 459; Phys. Lett. B 302 (1993) 396}
 \lref\opena{I. Kostov, Field theory of open and closed strings with
discrete
target space, Saclay preprint  SPhT/94-097, to be published in \pl }
\lref\shiba{N. Ishibashi and H. Kawai, \pl B 322 (1994) 67}
 \lref\ike{M. Ikehara, N. Ishibashi, H. Kawai, T. Mogami, R. Nakayama and
N. Sasakura,
 KEK preprints  KEK-TH-402,  and   KEK-TH-411, August 1994}
\lref\kkw{H. Kawai, N. Kawamoto, T. Mogami and Y. Watabiki, \pl B 306
(1993) 19,
N. Ishibashi and H. Kawai \pl  B 314 (1993) 190
}
\lref\franc{F. David, \np B 368 (1992) 671}
\lref\moga{Tsuguo Mogami, KEK preprint KEK-TH-426}
\Title{}{\vbox{\centerline{
    LOOP SPACE HAMILTONIAN }
\vskip4pt\centerline{
  FOR  $c \le 1$   OPEN   STRINGS  }
}}
 \bigskip\centerline{ I. K. Kostov \footnote{$^\ast $}
{ on leave from the Institute for Nuclear Research and Nuclear Energy,
72 Boulevard Tsarigradsko Chauss\'ee, 1784 Sofia, Bulgaria}}
\bigskip\centerline{{\it Service de
 Physique Th\'eorique}  \footnote{$^{\dagger}$}{
Laboratoire de la Direction
 des Sciences de la Mati\`ere du
Commissariat \`a l'Energie Atomique}{\it  de Saclay} }
\centerline{{\it CE-Saclay, F-91191 Gif-sur-Yvette, France}}

\vskip .3in

\baselineskip8pt{We construct a string field Hamiltonian describing the
dynamics of open and closed strings with  effective target-space dimension
$c\le 1 $. In order to do so, we first derive the Dyson-Schwinger equations
for the underlying large $N$ vector+matrix model and formulate them as a set of
constraints satisfying decoupled Virasoro and U(1) current algebras.  The
Hamiltonian consists of a bulk and a boundary term having different scaling
dimensions. The time parameters corresponding to the two terms are interpreted
from the the point of view of the fractal geometry of the world surface.
 }

\bigskip
\bigskip
\bigskip
\leftline{Submitted for publication to: {\sl Physics Letters B}}%
\rightline{ SPhT/95-001}
 \Date{01/95}
\baselineskip=20pt plus 2pt minus 2pt

\newsec{Introduction}

   The string field theory  seems to be the most complete and
geometrically the most natural
formalism  for studying strings in interaction. The construction of such a
theory  implies a
decomposition of the  world sheet into propagators and vertices which is a
subtle and highly
nontrivial problem and can be done in different ways. An unambiguous
prescription for construction
 of a string field theory stems from the discretization  of the   string
path integral in which both
the world sheet and the target space are discretized \Idis .
   Recently, the strings with discrete target space
were reformulated  in terms of a transfer matrix formalism \kkw \shiba
\ike. One of the interesting
aspects of this approach is the interpretation of the   time parameter for
the string field
Hamiltonian    as  geodesic distance on the world sheet of the
string\foot{In
an equivalent approach
based on stochastic quantization \jjev , the interpretation of this
parameter is as a
fictitious stochastic time.}.  If this interpretation is correct, then the
dimension of the  time parameter  is related to
the "intrinsic fractal dimension" of the world surface.
The computation of the latter in the Liouville theory formalism  is a very
difficult and still unsolved problem  \franc . Moreover, the finite-time
correlation functions can be interpreted as correlations between points at
given geodesic distance on the world sheet.

In this letter we resolve some minor inconsistencies of the transfer matrix
approach of ref. \shiba\ and  extend it to the case of interacting closed and
open  strings. Our motivation is the hope that such an extension would help to
understand the fractal structure of random surfaces with boundaries.
 Our starting point will be the
  soluble large $N$ vector+matrix model  proposed in ref. \opena .
  We will derive the
loop space Dyson-Schwinger equations for this model  and reformulate them as  a
set of  decoupled Virasoro and $U(1)$
current algebra constraints, one for each point $x$ of the target space.
The Virasoro constraints can
be obtained from a  bulk Hamiltonian $\CH$ which generalizes the closed
string  Hamiltonian
constructed in \shiba . Its time parameter $t$  measures the geodesic
distance on the world sheet.
The $U(1)$ constraints follow from   a boundary Hamiltonian $\CH_B$ whose
evolution  parameter $t_B$  allows an interpretation as the geodesic distance
along  the edge of the world
surface of the open string. We will evaluate the dimensions of these two
time parameters and
interpret their meaning  from the point of view of the fractal structure
of the world surface.

 \newsec{ Large $N$ field theory  for discretized open and closed
strings}

By a string we understand an oriented  chain of particles immersed in the
target  space $\Z$ and such that the coordinates of the adjacent particles
either coincide or differ by   1. The configuration space of strings  consists
of all closed and  open paths $\Gamma = [x_0x_1x_2...x_n]; \ x_k\in\Z,
x_{k+1}-x_k=0,\pm 1$.   The evolution of the string can be decomposed into
elementary moves  creating or annihilating a  particle
$([...xx'...]\leftrightarrow [...xx"x'...])$ or changing its topology
$([...x...x...]\leftrightarrow [...x[x...x]x...])$.
It is evident that the world surface of such a string will represent a
triangulated surface immersed in $\Z$.
  A string theory with    effective dimension
$c=1-6/m(m+1)  $ of the target space  will be obtained by modifying the
vacuum state   by means of a   background momentum $p_0=1/m$.

The discretized   string field theory  describes  the perturbative
expansion of a one-dimensional model of vector and matrix fields with the same
color  structure as the
  $U(N)$ lattice gauge theory with quarks
\opena . The   $N$-vector "quark" field $\psi_x ,\psi_x^*$ is associated
with the points $x \in  \Z $ and creates the ends of the open string.
 The line elements of the discretized string are represented by an  $N\times N$
matrix  "gluon" field $A_{xx'} $   defined for the  oriented pairs
of points $\{x, x'\}$ such that $|x-x'|\le 1$.      The diagonal matrices
$A_{xx}$ are hermitian and the  nondiagonal ones satisfy the
condition $A_{xx'}=A_{x'x}^{\dag}$.
 The gauge invariant operators creating closed and open strings are  the
the Wilson   lines and
loops
 \eqn\lllp{\Omega_{[x_0x_1...x_n]}= \psi^*_{x_0}
A_{x_0x_1}A_{x_1x_2}...A_{x_{n-1}x_n}\psi_{x_n}}
  \eqn\wllp{W_{[x_0x_1...x_nx_0]}= \tr \
(A_{x_0x_1}A_{x_1x_2}...A_{x_nx_0}).}
The partition
function is defined  as
\eqn\partf{\CZ=\int  dA \ d\psi d \psi^*
e^{\CW [A, \psi ,  \psi^*]}}
\eqn\actn{\eqalign{
\CW [A, \psi ,  \psi^*] = &-\half   \tr  \sum_{x,x'}
   A _{xx'}A _{x'x }
+{\lambda \over 3\sqrt{N}  }   \tr \sum_{x,x',x"}
A_{xx'}A_{x'x"}A_{x"x}\cr
   &+\sum_{x,x'}   \psi _x^{*}(- \delta_{xx'} + {\lambda_{B} \over
\sqrt{N}}
  A_{xx'})\psi_{x'} .\cr }  }
 where the   $\lambda$ is the  bare  string
tension and $\lambda_B$   is the bare
"quark" mass. The components of the $\psi$-field can be considered as
commuting or anticommuting
variables. The   two choices  lead to  different  signs of the string
interaction
constant.

One can write a collective-field Hamiltonian in terms of the
gauge-invariant
loop fields \wllp -\lllp\ following the general recipe \djv . It is
however possible   to find a
Hamiltonian acting in a smaller configuration space, namely, the
contours   associated with a
single point $x$. The corresponding Wilson loops and lines involve only
the diagonal matrices
$A_{xx}$. Indeed, the action  \actn\  is quadratic with respect to   the
nondiagonal matrices
$A_{xx'}= A^{\dagger}_{x'x}, \  x'=x\pm 1 , $ and the corresponding
integration is trivial.
   Therefore, if we restrict
ourselves to contours localized at a single point $x$,   the  model \partf
-\actn\  can be formulated
only in terms of the "quark" fields $\psi_x, \psi^*_x$ and the hermitian
matrix field $M_x$
\eqn\tfkk{  M_x^{ij}  = A_{xx}^{ij}-{ \sqrt{N}\over   2\lambda}
\delta^{ij}.}
 The reduced Hilbert space is spanned by the operators
\eqn\akppo{  W_{x}(L)   =   \tr \  e^{L  M_x } ,\ \ \  \Omega_{x}(L)
=\psi_{x}^{*} \
e^{L  M_x } \ \psi_{x}  }
or their Laplace transforms
  \eqn\tfdr{ \tilde W_{x}(Z)= \int_0^{\infty} dL \ e^{-L Z}   W_x(L)
= \tr \ {1\over Z- M_x}  }
 \eqn\tfdrr{ \tilde \Omega_{x}(Z)  =  \int_0^{\infty} dL \ e^{-LZ }
\Omega_x(L)
 =  \psi_{x}^{*}  {1\over Z- M_{ x}} \ \psi_{x}.
}

The generating functional for the correlation functions of these operators
is obtained from \partf\
by introducing  the  source terms
 \eqn\srrs{ J\cdot W= \sum_x \int _{0}^{\infty} dL J_x(L)W_x(L)=
\sum_x\oint {dZ\over 2\pi i}\tilde  J_x(Z) \tilde W_x(Z)}
\eqn\srrso{ J^B\cdot\Omega = \sum_x \int _{0}^{\infty} dL
J_x^B(L)\Omega_x(L)=
\sum_x\oint {dZ\over 2\pi i} \tilde J_x^B(Z) \tilde \Omega_x(Z)}

Integrating over the nondiagonal $A$-matrices
 and  adding the  source terms, we find the following expression for
the    generating
functional for the loop  fields
\eqn\mtra{\CZ[J,J^B]=\int \prod _{x}\ d\psi_x d  \psi_x^* d M_x
\ \ e^{J\cdot W+J_B\cdot\Omega+ \CA [W, \Omega]}}
 \eqn\mtrb{   \CA [W, \Omega]
=\half  \sum_{x,x'}  C_{xx'}  \int _0^{\infty} d L\Big[{1\over L}   W_x(L)
W_{x'} (L)
+ \Omega_x(L)\Omega_{x'} (L \Big]}
  where $C_{xx'}$ denotes  the adjacency matrix of the target space
\eqn\adj{C_{xx'} = \delta_{x,x'+1}+\delta_{x,x'-1}.}
 We have  eliminated the explicit dependence on the   couplings $\lambda$
and $\lambda_B$
by   shifting the sources and rescaling the $\psi$-fields.
  The sources now should be considered as small perturbations around the
polynomial potentials
$  V(Z)$ and $V^B(Z)$.

The action \actn\  leads to   $V(Z) =$ [polynomial of third degree] and
$V^B=$ [constant].
However one can consider   more general polynomial potentials. The
multicritical  regimes of the closed string
are obtained by tuning the potential $V$.

The effective action \mtrb\ has an  evident geometrical interpretation. The
integration over the nondiagonal $A$-matrices organizes the elementary
triangles that compose the
world sheet into rings (the first term) and strips (the second term).  A
ring can be interpreted as
the   amplitude for   a
    jump of a closed string with length $L$ from the point $x$ to the
adjacent
point $x'=x\pm 1$. Similarly, a strip  represents a jump $x\to x'$ of an
open string with
length $L$.     In this way the  evolution of the string string can  be
decomposed into
 elementary processes representing  either  splittings or  joinings,  or
propagation, but not both.
  This factorization is an important feature  of the strings with discrete
target space and gives
the clue for the exact solution of the  theory.

  The sum over the embeddings of the world surface  can be thought of as
the sum
over all possible configurations of the domain walls   separating the
domains with constant $x$ (see fig.7 of ref. \kko). This is the
partition function of a gas of
nonintersecting loops and lines on the world surface.   Each
loop or line has two orientations and therefore has to be taken with a
factor of  2. The endpoints of the
domain lines are arranged along the   boundary of the world surface formed
by the ends of open
strings.
 In ref. \kko\ this type of boundary is
called "Neumann boundary" because it is characterized by a free boundary
condition.
On the other hand, the  pieces of
the boundary associated with the incoming and outgoing loops are
characterized by a constant
(Dirichlet) boundary condition.  The domain walls do not touch the
Dirichlet boundary.   Once we have  restricted the Hilbert space to the
operators of the form   \akppo ,    the "time"
direction on the world sheet is fixed: the lines of constant time go along
the domain walls.

 The model \mtra - \mtrb\ describes strings whose target space is the
one-dimensional lattice
$\Z$ characterized by the adjacency matrix  \adj .
 The dimension of the target space, or the central charge of the matter in the
language of 2d quantum gravity, can be   lowered to $c<1$
 by introducing   a background momentum $p_0$ coupled to the intrinsic
curvature of the world sheet.
 This is explained in the context of the loop-gas model in ref. \kko .
In  our  formalism  this is achieved by using twisted adjacency  matrix \adj
 \eqn\ctrm{ C_{xx'} \to
\cases{C_{xx'}^{(p_0)}=  e^{ip_0} \delta_{x,x+1}
+e^{-ip_0}\delta_{x,x'-1}
,& for closed strings ;\cr
  C_{xx'}^{(  p_0/2)}=  e^{ip_0/2} \delta_{x,x+1} + e^{-ip_0/2}
\delta_{x,x'-1}  ,& for open strings  \cr } }
  For example, the   action \mtrb\  becomes
  \eqn\mtrbb{   \CA [W, \Omega]
=\half  \sum_{x,x'}    \int _0^{\infty} dL\Big[C_{xx'}^{(p_0)} {1\over
L}   W_x(L) W_{x'} (L) +
C_{xx'}^{(  p_0/2)} \Omega_x(L)\Omega_{x'} (L)\Big].}
  In the loop gas picture  the twisted ajacency matrix means that the domain
walls are weighed by  phase factors depending on their orientation.

The background momentum changes the vacuum of
the string theory and the  effective dimension of the
target space then  becomes
 \eqn\cchr{c=1-6 {( g - 1)^2 \over g}, g= p_0+1. }
The momentum space is
periodic with period $2$, but nevertheless it makes  sense to consider the
parameter $g$ in the interval $0<g<\infty$. The integral part of $g$
specifies the critical regime
obtained by tuning the potential $V(Z)$.   The dense phase of the loop gas is
described by
the interval $0<g<1$, the dilute phase   by   the
interval $1<g<2$, and the $m$-critical regime  by $m-1<g<m$.  The
half-integer values $g= m-\half,
m=1,2,...,$ describe the possible critical regimes of strings without
embedding space.

In what follows by adjacency matrix $C_{xx'}$ we will understand the twisted
adjacency matrix \ctrm .

 \newsec{Loop equations }

 The invariance of the integral \mtra\
 under an  infinitesimal  change   of
variables
\eqn\chv{  M_{ x}\to M_{ x}+{\epsilon_{ x}\over Z- M_{ x}}
;  \  \  \psi_{ x}\to (1+  {\theta  _{ x} \over Z-  M_{ x} }   ) \psi_{
x}  }
 yields  a closed set of
    Dyson-Schwinger equations  for the loop fields \akppo
\eqn\lopl{\eqalign{ &\langle \int_0^{L}dL '  W_x(L ')   W_x( L-L')\ + \
\int_0^{\infty} dL ' [L '   J_x(L ') + \sum_{x'}C_{xx'}  W_{x'}(L ')]
 W_x(L +L ')
\cr &+   \int_0^{\infty} dL ' L '  [  J_x^B(L ') +
\sum_{x'}C_{xx'}  \Omega_{x'}(L ')]   \Omega_x(L +L') \rangle =0\cr}}
\eqn\loplo{ \langle  \    W_x(L )+ \int_0^{\infty} dL ' [  J_x^B(L ') +
\sum_{x'}C_{xx'} \Omega_{x'} (L')]   \Omega_x(L +L ')
  \rangle =0\ \ \ \ \ \ \ }
where we denoted by $\langle \ \ \rangle$  the expectation value with the
measure   \mtra  .
 The normalization of the closed string field corresponds to
 \eqn\nnor{  \langle W_x(0)\rangle =N .}

  We can alternatively formulate the loop equations in terms of the Laplace
transformed loop fields
\tfdr\ and \tfdrr . The equations read
\eqn\uhh{\eqalign{ &\langle \tilde W_x^2(Z)+{1\over 2\pi i} \oint dZ'\
{\tilde W_x(Z')
 \over Z-Z'}\ [\p   \tilde J_x(Z')+\sum _{x'}C_{xx'}\tilde W_{x'}(-Z')]\cr
&+   {1\over 2\pi i} \oint dZ' \ {
\tilde \Omega_x(Z') \over(Z-Z' )^2} \  [\tilde J_x^B(Z')+\sum_{x'}
C_{xx'}\tilde
\Omega_{x'}(-Z')]\rangle =0 \cr}}
\eqn\oror{\langle \tilde W_x(Z)+{1\over 2\pi i}\oint dZ'\ { \tilde
\Omega_x(Z')   \over Z-Z' } \
[\tilde J_x^B(Z')+\sum_{x'}
 C_{xx'}\tilde \Omega_{x'}(-Z')]\rangle =0.}
The contour of integration encloses   the singularities  of $\tilde
W_x(Z')$ and
 leaves outside  the   point $Z'=Z$  as well as the singularities
of $\tilde W_x(-Z')$.
It is selfconsistent to assume that in any finite order of the $1/N$ expansion
 the correlation functions of   $W_x(Z)$ and $\Omega_x(Z)$  are
meromorphic  functions defined in the  $Z$ plane  cut along the same interval
$X<Z<Y$.   The location of the endpoints can be
determined as usual by the
condition
 $\langle\tilde  W_x(Z)\rangle  = N/Z + \CO(1/Z^2), \ \ Z\to\infty$. The
contour integrals in \uhh
-\oror\ makes sense only if the  intervals $[X,Y]$ and $[-Y,-X]$ do not
overlap, i.e., if $Y<0$.  The
typical length of  strings is $L \sim |Y|^{-1}$ and the  critical
point  is   achieved when    $Y\to
Y^*=0$.

   \newsec{The scaling limit }

    The typical world surface
is characterized by its   area $A $, the length $L $ of its  Dirichlet
boundaries and the
length $
L_B$ of its  Neumann boundaries.
Near the critical point $\lambda = \lambda^*,  \lambda_B =
\lambda^*_B$  these  parameters  diverge as
 \eqn\aaaa{ A\sim {1\over \lambda^*-\lambda}, \
L\sim - {1\over Y}  ,\
  L_B\sim {1\over  \lambda^*_B- \lambda_B}.}

Let us  introduce a scale parameter  (elementary length)  $a$   by
 \eqn\ctff{ Y = -Ma.}
 and rescaling the   variables as
\eqn\reesc{L =\l /a , \ \ Z = z a.}
The scaling limit is characterized by the   renormalized
parameters (the string interaction constant $\kappa$, the string tension
$\Lambda$ and the
  "quark" mass $\mu$)   defined by
\eqn\renop{\kappa ={1\over  a^{ 1+g}N} , \ \lambda ^* -\lambda \sim
a^{2\nu}\Lambda, \ \
\lambda _B^*-\lambda_B \sim a^g\mu}
where
\eqn\nunu{\nu  =\cases{g,& if $ g<1$ ;\cr 1,&if $ g>1$ \cr } }
The parameter $M$  is a function of  $\Lambda$; in an appropriate
normalization it is given by
\eqn\gghbm{M^{2\nu} = c^2\ \Lambda, \ \ c ^{\nu/g}= 2\sin  (\half \pi g).}

The   so called "string susceptibility" exponent is equal to
$\gamma_{{\rm str}}= -|g-1|/\nu$.  For given $n=-2\cos \pi g$ the branch
$0<g<1$
describes the dense phase
of the loop gas, the
branch $1<g<2$  describes the dilute  phase, and the branches with $g>2$
correspond to an
infinite sequence of multicritical phases.    The
scaling exponents follow from the analysis of the loop equations \uhh\ and
\oror  (see \Idis, Sect.
3).

 As
mentioned above, we assume polynomial potentials $\tilde J_x(z) =- V(z),
\tilde J^B_x(z)= -V^B(z)$.
By shifting the loop fields by appropriate  polynomials in $z$,   the
dependence of the l.h.s. of the
loop equations on the potential can be eliminated. The r.h.s. of the loop
equations then is no more
zero, but a polynomial in $z$.
The   renormalized   loop fields and sources are defined by
\eqn\rnlp{\tilde  W_x(Z) =    P (Z)+{1\over a\kappa} \tilde w_x(z), \ \
\tilde\Omega_x(Z)=   P_B(Z)+
{1\over\sqrt{ a\kappa}}\tilde \omega _x(z)}
\eqn\soorts{ \tilde J_x(Z)=-V(Z)+\kappa\tilde j_x(z), \ \ \tilde
J_x(Z)=-V^B(Z)+\sqrt{\kappa}
\tilde j_x^B(z)}
where the polynomials $   P (Z)$ and $ P_B(Z)$ are chosen  to cancel the
dependence
on the potential  of the l.h.s. of the loop equations.

The   disc amplitudes ($\kappa =0$)
  \eqn\diisk{  \tilde w_c(z) = \langle \tilde w_x(z)\rangle, \
 \tilde \omega _c(z) = \langle\tilde \omega _x(z)\rangle }
do not depend on $x$ and satisfy the
following functional
equations \kko
 \eqn\stdv{
 \tilde w _c  ( z )   ^2 +\tilde w _c  ( -z )  ^2-  2\cos  \pi g \ \
\tilde w_c(z)   \tilde w _c(-z )  =  \Lambda^{   g/\nu}  }
\eqn\smmp{ 2 \sin \half \pi g\  \ \tilde \omega _c( z) \tilde\omega_c ( -z)
  +  \tilde w _c ( z ) + \tilde  w _c ( -z) = 2\mu  .}
     The   solution of
these equations in the case of a string with massless ends ($\mu =0)$ is
given by simple algebraic
functions
  \eqn\wwl{ \tilde w_c(z)= -{  (z+\sqrt{z^2-M^2})^g +(z
\sqrt{z^2-M^2})^g \over 2\cos   \pi g/2} }
 \eqn\ool{\tilde \omega _c(z)=-   {  (z+\sqrt{z^2-M^2})^{g/2}
+(z-\sqrt{z^2-M^2})^{g/2}
 \over  \sqrt{ 2\sin  \pi g/2}}
 .}
 The solution \wwl\ corresponds to the normalization \gghbm.

The amplitudes for $c=1$ are obtained by taking the limit $g\to 1$ of
these expressions; we see
that logarithmic singularities occur in this limit.    The complete
solution of \stdv\ is given
in ref \kko.

The renormalized loop fields satisfy the same loop equations as the bare
fields, but
a polynomial of  $z$ will appear on the r.h.s. of the equations \uhh\ and
\oror.  For generic value of the
parameter $g$ only the constant term of this polynomial will have the right
dimension,   but for
half-integer $g$ higher powers of $z$ can survive.
  For example, in the case $g=3/2$  the
r.h.s. of \uhh\ contains terms proportional to $z\Lambda $ and $z^3$.

\newsec{The algebra of constraints}

The  logarithm of the  partition function  $ \CZ[J, J^B]$   is
 the generating functional for the connected loop
correlators
   \eqn\core{ \langle w_x(\l)...\omega_{x'}(\l')...\rangle = \Bigg(
{    \delta\over\delta j_x(\l)} ...
{\delta\over\delta j_{x'}^B(\l')}...\ln
Z[j, j^B] \Bigg)_{j=j^B=0} .}

The loop equations \lopl\ and \loplo\  can be formulated as
second-order variational
equations for    $ Z[j, j^B]$
\eqn\viras{\CT_x(\l)\ \CZ =   \CB_x(\l) \ \CZ\  =0}
where
\eqn\oplc{\eqalign{
\CT_x(\l)=&  \int_0^{\l} d\l' {\delta \over \delta j_x(\l') } {\delta
\over \delta j_x(\l -
\l') } \cr
&+ \int_0^{\infty}d\l'
     \sum_{ x'}C_{xx'}\Bigg({\delta \over \delta j_{x'}( \l') }
{\delta \over \delta j_x(\l+\l') }+ \kappa  \l'   {\delta \over \delta
j_{x'}^B( \l') }
 {\delta \over \delta j_x^B(\l+\l') } \Bigg)
\cr
&  + \kappa^2  \int_0^{\infty}d\l'\  \l' \ \Big(   j_{x}( \l') {\delta
\over \delta j_x(\l+\l') }
+      j^B_{x}( \l') {\delta \over \delta j_x^B(\l+\l') }\Big) \cr
 }}
\eqn\oplo{\CB_x (\l)=
{\delta \over \delta j_x(\l)}
+\int_0^{\infty}d\l'\Bigg(\kappa  j^B_{x}( \l')
 {\delta \over \delta j_x^B(\l+\l') }+
   \sum_{x'}C_{xx'}
{\delta \over \delta j_{x'}^B(\l')} {\delta \over \delta j_x^B(\l+\l') }
 \Bigg) }
A consistency condition for the integrability of these equations is that
the
  operators $\CT,\CB $ should close a Lie-algebra. This is indeed the
case:
 \eqn\commm{\eqalign{
&[\CT_x(\l ), \CT_{x'}(\l ')]=\kappa^2 (\l-\l')\delta_{xx'}\ \CT_x(\l+\l'
);\  \ \
[\CB_x(\l ), \CB_{x'}(\l ')]=0\cr
&[\CB_x(\l ), \CT_{x'}(\l ')]= \kappa^2\l\ \delta_{xx'}\ \CB_x(\l+\l' )\cr}
 }
The constraints  form a set of decoupled algebras, one for each point $x$.
Each  such
algebra is a semi-direct product of a "continuum" Virasoro and an $U(1)$
current algebra, truncated
to $\l >0$. The fact that the closed string sector is described by
decoupled Virasoro algebras has
been already pointed out by Ishibashi and Kawai in \shiba .

\newsec{The loop space Hamiltonian}

The   loop space Hamiltonian  will be constructed following the same logic
as in
ref.\shiba.    Let us introduce a representation of the fields $W$ and
$\Omega$ as  creation and
annihilation operators of the string states. Denote by $\Psi_x(\l)$ (
$\Psi_x^{\dag}(\l)$) the
operator that annihilates (creates) an open string state with length $\l$
located at the point $x$.
 Similarly we define the  operators $\Phi_x^{\dag} (\l)$ creating a closed
string with a
marked point and the corresponding annihilation operator $\Phi_x(\l)$.
These operators satisfy the canonical commutation relations
\eqn\ccr{[\Psi_x(\l),\Psi_{x'}^{\dag}(\l')]= \delta_{xx'} \delta (\l- \l
'), \ \
[\Phi_x(\l),\Phi_{x'}^{\dag}(\l')]=\delta_{xx'} \delta (\l-\l ').}
The vacuum state is defined by
\eqn\vacs{\langle 0|\Psi_x^{\dag}(\l)=\langle 0 |\Phi_x^{\dag}(\l)=0,  \ \
\
\Psi_x (\l)|0\rangle =  \Phi_x (\l)|0\rangle= 0.}

Let us introduce   the shorthand notations
\eqn\connv{\eqalign{\{f, g| h\}&=\int_0^{\infty} d\l d\l' f(\l) g(\l')
h(\l +\l'),\cr
\{h|f, g\}&=\int_0^{\infty} d\l d\l'  h(\l +\l') f(\l) g(\l')\cr
f\cdot g &= \sum_x\int _0^{\infty} d\l f_x(\l)g_x(\l) \cr}}
and define the two  Hamiltonians
\eqn\hamul{
\eqalign{
\CH   =  & \sum_x \Big(  \{ \Phi_x^{\dag},  \Phi_x^{\dag}| \l \Phi_x \}
+  \kappa^2\{\Phi_x^{\dag} |  \l \Phi_x  , \l \Phi_x \}
+\kappa^2\{  \Psi_x^{\dag}  |  \l \Psi_x   ,
\l \Phi_x, \} \Big)\cr
+&\sum_{x, x'} C_{xx'}  \Big(   \{\Phi_x^{\dag}|  \Phi _{x'}^{\dag}, \l
\Phi_x \}+ \kappa
\{ \Psi_x^{\dag}  |   \l \Psi _{x'}^{\dag} ,
\l \Phi_x, \}\Big) \cr}
}
\eqn\hamb{ \CH_B  =    \sum_{x, x'} C_{xx'}   \{ \Psi_x^{\dag} | \Psi
_{x'}^{\dag} ,\Psi_x \}+
   \kappa \sum_x
 \{ \Psi_x^{\dag} |   \Psi_x  ,\Psi_x \} + \Phi^{\dag}\cdot\Psi
}

We will argue that the limit $t, t_B \to\infty$ of   the functional
\eqn\ppff{
 \CZ_{t,t_B} [J,J^B]=
\langle 0|e^{ w_c\cdot \Phi +\omega_c \cdot \Psi }   e^{-t\CH
-t_B\CH _B } e^{ j \cdot    \Phi ^{\dag}
+  j ^B \cdot  \Psi  ^{\dag}} |0\rangle  }
  gives a
  formal solution to the constraints \viras .
First let us note that in the classical limit $\kappa=0$,  the functional
\ppff\ does not depend on
the two time parameters $t$ and $t_B$ and is given by its value at
$t=t_B=0$
\eqn\lalala{\Bigg(\CZ_{t,t_B} [j,j^B]\Bigg)_{\kappa =0} = e^{w_c\cdot j  +
\omega_c\cdot j^B }.}
 This can be checked by taking the derivatives in   $t$ and $t_B$ and using
the fact that   $w_c$ and $\omega_c$ satisfy the planar loop equations.
    The solution for $\kappa \ne 0$ satisfies the differential equations
 \eqn\ohhw{{\p \over \p t} \CZ_{t,t_B} [j,j^B] = \sum_x \int _0^{\infty}
d\l \ \l\  j_x(\l)\
\CT_x(\l) \ \CZ_{t,t_B} [j,j^B]}
\eqn\ohhwo{{\p \over \p t_B} \CZ_{t,t_B} [J,J^B] = \sum_x \int _0^{\infty}
d\l \  j_x^B(\l)\
\CB_x(\l) \ \CZ_{t,t_B} [j,j^B].}
One can think of the evolution of the string as a sequence of elementary
processes
of splitting and joining. The operators \hamul\ and \hamb\   describe the
evolution with respect to  two   time  parameters $t$ and $t_B$.
We  can consider only one time $t$ and introduce a
dimensional constant $R$ such
 that $t_B = R t$.   The
parameter $R$ will compensate the difference of the time scales in the bulk
and at the boundaries.
 The bulk Hamiltonian $\CH$ involves    processes that  can occur at any
point of a string   (see ref. \shiba \ for a description). This is the
origin of the $\l$-factors
associated with the annihilation operators. On the other hand, the
boundary Hamiltonian $\CH_B$
describes elementary processes that  are possible only at the   endpoints
of an open string. After an
infinite amount of "time" the system comes to an equilibrium and does not
evolve any more.
It follows from \ohhw\ and \ohhwo\ that the functional
\eqn\nnan{\CZ[J,J^B]=\lim _{t' t_B\to\infty}  \CZ_{t,t_B} [J,J^B]}
satisfies the constraints \viras .

An important feature of the   Hamiltonians \hamul\ and \hamb\ is that they
do not contain
  tadpole terms, in contrast with the Hamiltonian considered in \kkw\ and
\ike.  Such terms can appear
only for half-integer values of $g$ corresponding to the multicritical
regimes of a random surface
with $C=0$.  Of course, the tadpole terms are present far from the
critical point but
they are multiplied by positive powers of the cutoff $a$ and disappear in
the scaling limit.
This means that the dynamics in the scaling limit is played at large
distances and the fraction of
the strings with infinitesimal length is always small. This fact  can be
explained with the processes describing propagation of strings, which  increase
the length of the initial string state.  Therefore, in order to obtain a
nontrivial expectation value, we  multiplied   the left vacuum by $ e^{w_c\cdot
j  +
\omega_c\cdot j^B }$.
Let us also remark that the bulk and boundary  Hamiltonians do not commute.
Therefore one can  construct more complicated Hamiltonians    by adding
commutator terms. In this way one can search a connection with the Hamiltonian
for open $C=0$ strings recently  proposed in \moga .

 The dimensions of the fields and parameters, in terms of
units of length
 $L_0$, are
\eqn\dimss{[\Phi_x(\l)]=L_0^{ g}, \Phi_x^{\dag}
(\l)]=L_0^{-g-1},[\Psi_x(\l)]=L_0^{   g/2},
[\Psi_x^{\dag} (\l)]=L_0^{- g/2-1}}
\eqn\dimrs{[\l] =L_0, [M]=L_0^{-1}, [\kappa]=L_0^{-g-1},
[\Lambda]=L_0^{-2\nu}, [\mu]=L_0^{-g}. }
This gives the following values for the dimensions of the two time
parameters
\eqn\dimt{[t]=L_0^{ g-1},\ \  [t_b]=L_0^{g/2}.}

The parameter $t$ was identified in \kkw\ with the time direction for  a
special choice of the
coordinates on the world sheet, the so called "temporal gauge". A
peculiarity of this gauge choice
is that the string is allowed to suffer unlimited number of splittings
before disappearing into the
vacuum. This makes a sharp contrast with the conformal gauge. The
geometrical meaning of the
  time parameter $t$ is the    geodesic (minimal) distance on the world
sheet. More concretely,
the time coordinate of a point close to the time slice $t_0$ is
$t_0+\delta t$, where $\delta t$
is the minimal distance from this point to the time slice $t=t_0$.

In our case, the interpretation of the time  $t$  as geodesic distance on
the world sheet makes
sense only if a statistical interpretation of the sum of the surfaces is
possible, i.e., if all
Boltzmann factors are positive.  This  condition is fulfilled  if $g<2$
and  $n=-2\cos \pi g>0$. We
also exclude the dense phase $g<1$ in which the area between the domain
walls vanishes and  the dimension of
the time $t$ is negative.  This leads us to
 the interval $1\le g\le 3/2$.    In this  interval, which describes string
theories with $0\le c\le 1$,
 the intrinsic fractal dimension $d_H$ of the
world sheet is defined  by $ A(r) \sim r^{d_H}$ where $A(r)$ is  the area
of a circle with radius
$r$. The dimension if the area is $[A]=[1/\Lambda] =L_0^{2\nu}$ where $\nu
=1$.  Hence, $d_H= 2
/(g-1)$ or, in terms of   $c$,
\eqn\dffd{d_H(c)  = {24\over
1-c+\sqrt{(1-c)(25-c)}}, \ \ \ 0\le c<1 . }
 In particular, for the unitary series
$c=1-6/m(m+1)$  this formula gives the  integer  value $d_H=2m$
 obtained  in \ike , where a different Hamiltonian was used.  We see that
 the intrinsic  fractal dimension   of the
world sheet increases with $c$ and becomes infinite  at the critical
dimension  $c=1$.   The fractal dimension
$d_H =4$ for strings without embedding ($C=0$) has been predicted both by
numerical simulations
\ref\tsuda{Tsuda and Yukawa, \pl B 305 (1993) 223} and analytic arguments
\ref\wawa{J. Ambjorn and Y.
Watabiki, NBI preprint NBI-HE-95-01, January 1995}.  It is also evident
that eq.   \dffd\  is  not
true when considered  for $C<0$ since the expected behavior of   of $d_H$
is $D=2$ in the "classical"
limit  $C\to -\infty$.  The numerical simulations give the bound $d_H
>3.5$ for $C=-2$ \ref\kazkaw{N.
Kawamoto, V. Kazakov, Y. Saeki and Y. Watabiki,
 {\it Phys. Rev. Lett.} 68 (1992) 2113}.

 The parameter $t_B$ can be interpreted, in a similar manner,
as the geodesic distance
along the boundary   representing a "quark" world line.
 The typical lengths of the Neumann and
Dirichlet boundaries are $\l\sim 1/M$ and $\l_B \sim 1/\mu$, correspondingly.
Note that
the  lengths measured along
the two types of boundary have  the same  dimension only for  $c=1$.

The  intrinsic fractal dimension
$d_H^B$ of the  Neumann boundary is defined by $\l_B (r) \sim r^{
d_H^B}$
where $\tilde\l (r) $
is  the  length of the   set of points of the boundary at geodesic
distance $d< r$ from given
point.   The interpretation of $t_B$ as geodesic distance gives the value
\eqn\bfd{  d_H^B= 2.}
This means  that for an observer living on the world surface,  the  true
linear extension of the
(Neumann) boundary will be the square root of its length.
Therefore, independently of the dimension of the embedding space, the
boundary will look from
 the interior of the surface as a random walk.

 As a future development one may think of calculating the correlation
functions of local operators
at   fixed geodesic distance by considering finite time intervals. Such
quantities have been
considered in \wawa .
 Finally, let us remind
 that the   closed strings with $C=1-6/m(m+1)$ can
be obtained from two
different  Hamiltonians. The Hamiltonian constructed in \shiba\ generates
$m-1$ decoupled Virasoro
algebras while this of   ref. \ike \ generates the $W_{m}$ algebra of
constraints on the
partition function. Each of the two approaches has its advantages and it would
be
helpful to understand better their relation.

The author   would like to thank Galen Sotkov and Y. Watabiki for  useful
discussions and F.
David for critical
reading of the manuscript.

\listrefs

\bye